\documentclass[prl,reprint,twocolumn,superscriptaddress,showpacs,floatfix]{revtex4-1}
\usepackage{amsmath}
\usepackage{mathrsfs}
\usepackage{txfonts}
\usepackage{amssymb}
\usepackage{graphicx}
\usepackage{hyperref}
\usepackage{ulem}
 \usepackage{overpic}
 \usepackage{psfrag}
\usepackage{tabularx}
\usepackage{diagbox}
\usepackage{array}
\usepackage{placeins}

\usepackage{bm}
\usepackage{dsfont}
\usepackage{mathtools}
\newcommand{\PreserveBackslash}[1]{\let\temp=\\#1\let\\=\temp}

\makeatletter

\begin{document}
\title{Exact matrix product state representations for a type of scale-invariant states}

\author{Huan-Qiang Zhou}
\affiliation{Centre for Modern Physics, Chongqing University, Chongqing 400044, The People's Republic of China}

\author{Qian-Qian Shi}
\affiliation{Centre for Modern Physics, Chongqing University, Chongqing 400044, The People's Republic of China}

\author{Ian P. McCulloch}
\affiliation{ Department of Physics, National Tsing Hua University, Hsinchu 30013, Taiwan}
\affiliation{Centre for Modern Physics, Chongqing University, Chongqing 400044, The People's Republic of China}

\begin{abstract}
Exact matrix product state representations for a type of scale-invariant states are presented, which describe highly degenerate ground states arising from spontaneous symmetry breaking with type-B Goldstone modes in one-dimensional quantum many-body systems. As a possible application, such a representation offers a convenient but powerful means for evaluating the norms of highly degenerate ground states. This in turn allows us to perform
a universal finite system-size scaling analysis of the entanglement entropy.  Moreover, this approach vividly explains why the entanglement entropy does not depend on what types of the boundary conditions are adopted, either periodic boundary conditions or open boundary conditions.  Illustrative examples include the ${\rm SU}(2)$ spin-$s$ Heisenberg ferromagnetic model, the ${\rm SU}(2s+1)$ ferromagnetic model, and the staggered ${\rm SU}(3)$ spin-1 ferromagnetic biquadratic model.
\end{abstract}
\maketitle

\section{Introduction}

Over the years, significant progress has been made in a proper classification of the Goldstone modes (GMs) arising from spontaneous symmetry breaking (SSB)~\cite{goldstone,Hnielsen, schafer, miransky, nambu, nicolis,  brauner-watanabe, watanabe, NG}. This leads to the introduction of type-A and type-B GMs~\cite{watanabe}.
Although many paradigmatic examples are known for SSB with type-A GMs, not much attention has been paid to SSB with type-B GMs. Recently, it has been found that
highly degenerate ground states,  a common feature for quantum many-body systems undergoing SSB with type-B GMs, are scale-invariant but not conformally invariant~\cite{FMGM,LLspin1,golden,SU4}. As a consequence, they constitute a counter-example for the speculation made by Polyakov~\cite{polyakov} that scale invariance  implies conformal invariance. Insights gained from this observation are certainly crucial in an attempt to achieve a complete classification of quantum phase transitions and quantum states of matter~\cite{entropy}.

The peculiarity of  quantum many-body  models undergoing SSB with type-B GMs is that they are exactly solvable, as far as highly degenerate ground states are concerned. This is due to an observation that they are described by the so-called frustration-free Hamiltonians~\cite{tasaki}. Actually, some of them are even integrable in the Yang-Baxter sense~\cite{FMGM,golden,SU4}. One of the recent developments is to demonstrate that highly degenerate ground states are subject to exact Schmidt decomposition, thus revealing self-similarities that reflect an abstract fractal underlying the ground state subspace~\cite{FMGM,LLspin1,golden,SU4}.
As a result, we are able to identify the fractal dimension with the number of type-B GMs. This may be achieved by performing a  finite system-size scaling analysis of the entanglement entropy~\cite{finitesize}, which in turn is evaluated from the norms of highly degenerate ground states.

Although efficient and powerful combinatorial methods are available to evaluate the norms in the representation-theoretic context, the presence of exact Schmidt decomposition for those highly degenerate ground states strongly indicates that they admit an exact matrix product state (MPS) representation. Here,
we emphasize that, although a MPS representation, as the simplest tensor network representation~\cite{1dTN,1d2dTN}, has been extensively exploited in numerical simulations of one-dimensional quantum many-body systems~\cite{1dTN},  an exact MPS representation is known {\it only} for a few physically meaningful quantum many-body systems, even if one restricts oneself to a ground-state wave function. In this regard,
the Affleck-Kennedy-Lieb-Tasaki (AKLT) state~\cite{AkLT} is a remarkable exception.  Hence, it is highly desirable to search for an exact MPS representation for highly degenerate ground states arising from SSB with type-B GMs in one-dimensional quantum many-body systems. Indeed, one might anticipate that an exact MPS representation for such a type of scale-invariant states offers a powerful means for extracting a variety of physical quantities, if it exists. In particular, it is convenient to evaluate the norms from an exact MPS representation for highly degenerate ground states.

In this work, we aim to address this intriguing question.  For our purpose, a matrix product operator (MPO) representation, developed in Ref.~\cite{yuping},  are adapted to represent lowering operators for a symmetry group. This makes it possible to develop a generic scheme to  evaluate the norms of highly degenerate ground states and perform
a universal finite system-size scaling analysis of the entanglement entropy.  Moreover, this approach vividly explains why the entanglement entropy does not depend on what boundary conditions are adopted, either periodic boundary conditions (PBCs) or open boundary conditions (OBCs).  Illustrative examples include the ${\rm SU}(2)$ spin-$s$ Heisenberg ferromagnetic model, the ${\rm SU}(2s+1)$ ferromagnetic model, and the staggered ${\rm SU}(3)$ spin-1 ferromagnetic biquadratic model.

\section{Generalities: Exact matrix product state representations for scale-invariant states}

Suppose a quantum many-body system, described by the Hamiltonian $\mathscr{H}$, undergoes SSB $G \rightarrow H$, with type-B GMs. Here, $G$ denotes a (semisimple) symmetry group and $H$ denotes a residual symmetry group. For simplicity, we focus on a one-dimensional quantum many-body system on a lattice, labeled by $j=1,\dots,L$, with $L$ being the system size.
As already demonstrated~\cite{FMGM,LLspin1,golden,SU4},  such a system admits highly degenerate ground states that are generated from the repeated action of a lowering operator of the symmetry group $G$ on the highest weight state or generalized highest weight states. As it turns out, the highly degenerate ground states are subject to exact Schmidt decomposition, which reveals self-similarities that reflect an abstract fractal underlying the ground state subspace~\cite{FMGM,LLspin1,golden,SU4}. That is, the highly degenerate ground states are scale-invariant, but not conformally invariant. Here and hereafter, we assume that no type-A GMs are involved. However, this is not restrictive, since type-A GMs do not survive quantum fluctuations in one dimension, according to the Mermin-Wagner-Coleman theorem~\cite{mwc,mwc2,GMA}.

To proceed, we have to distinguish two distinct situations, depending on what types of the boundary conditions, i.e., OBCs and PBCs, are adopted. We stress that the model Hamiltonian $\mathscr{H}$
commutes with $\tau$  under PBCs, where $\tau$ is the one-site translation operation. Hence, it is natural to expect that an exact MPS representation is translation-invariant, unless this discrete symmetry is spontaneously broken. Meanwhile, if it is spontaneously broken, then it is necessary to introduce a unit cell to accommodate the periodic structure of ground state wave functions.

\begin{figure}
	\includegraphics[angle=0,totalheight=4.5cm]{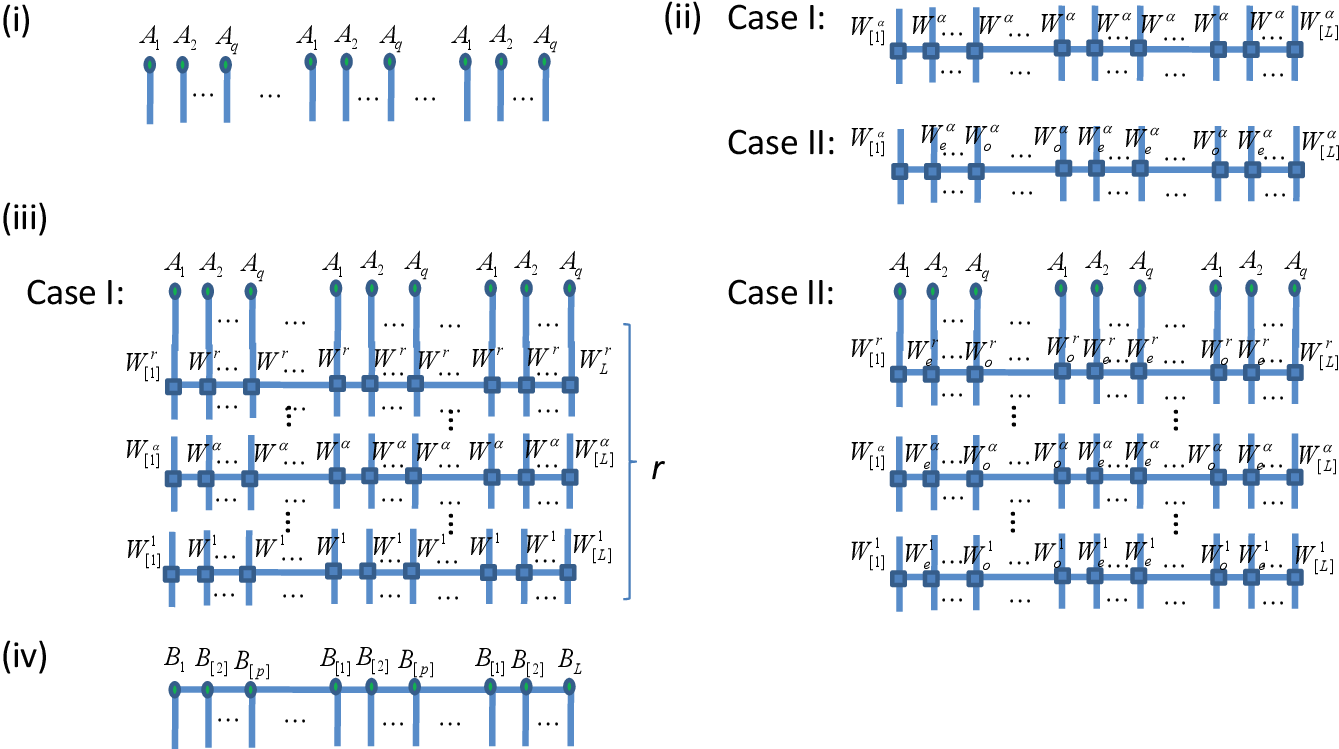}
	\caption{
		(i) A MPS representation for a factorized ground state $|\psi_0\rangle_q$, which denotes the highest weight state or a generalized highest weight state, with the period being $q$. It consists of $q$ complex numbers $A_1$, $A_2$, \ldots, $A_q$, if one does not take the physical index into account.  Here, $q$ is either 1 or an even integer.
		(ii) A MPO representation for $(F_\alpha)^{M_\alpha}$ ($\alpha=1$, 2, \ldots, $r$), consisting of two vectors $W_{[1]}^\alpha$ and $W_{[L]}^\alpha$, together with $L-2$ matrices $W_{[j]}^\alpha$, with $j=2$, \ldots, $L-1$. Here, $r$ is the rank of a (semisimple) symmetry group. We need to distinguish two cases:  (I) $W_{[j]}^\alpha$ ($j=2$, \ldots, $L-1$) is uniform, denoted as $W^\alpha$; (II) $W_{[j]}^\alpha$ is staggered, denoted as $W_e^\alpha$  for even $j$'s and $W_o^\alpha$ for odd $j$'s, respectively.
		(iii) A MPS representation for $|\psi(L,M_1,\ldots,M_r)\rangle_p=\prod_{\alpha=1}^r(F_\alpha)^{M_\alpha}|\psi_0\rangle_q$, with $|L,M_1,\ldots,M_r\rangle_p=1/Z_p(L,M_1,\ldots,M_r)|\Psi(L,M_1,\ldots,M_r)\rangle_p$ for case I and case II, respectively.
		(iv) A MPS representation for $|L,M_1,\ldots,M_r\rangle_p$, consisting of $p$ matrices $B_{[1]}$, $B_{[2]}$, \ldots, $B_{[p]}$, together with two vectors $B_1$ and $B_L$.
	We stress that the period $p$ is not necessarily identical to the period $q$, if the symmetry group is staggered.}
	\label{mps}
\end{figure}

If OBCs are adopted, an exact MPS representation for such a scale-invariant state follows from a prescription, consisting of the following three steps:
First, for a quantum many-body system, described by the Hamiltonian $\mathscr{H}$, with a given symmetry group $G$, one of degenerate ground states is the highest weight state, denoted as $|\psi_0\rangle$, or a generalized highest weight state, denoted as $|\psi_0\rangle_q$~\cite{FMGM,golden}. Here, $q$ denotes the period $q$, which is a positive integer. As a convention, we always keep the notation $|\psi_0\rangle$ to denote $|\psi_0\rangle_q$ if $q=1$. Usually, $|\psi_0\rangle_q$ is an unentangled factorized state, so it is in a MPS representation, with the bond dimension being one, as shown in Fig.~\ref{mps}(i). We remark that it consists of $q$ complex numbers $A_1$, $A_2$, \ldots, $A_q$, if one does not take the physical index into account.
Second, for a given (semisimple) symmetry group $G$ with the rank $r$, there are $r$ lowering operators, denoted as $F_\alpha=\sum_{j=1}^LF_{\alpha,j}$, with $\alpha=1$, 2, $\ldots$, $r$. For each of the $r$ lowering operators, its power may be turned into a MPO representation, as visualized in Fig.~\ref{mps}(ii).
Specifically, a generic MPO representation for $F_\alpha^{\,\,M_\alpha}$ may be written as follows
\begin{equation}
F_\alpha^{\,\,M_\alpha}= W_{[1]}^\alpha\cdots W_{[j]}^\alpha\cdots W_{[L]}^\alpha \,,
\end{equation}
where the two vectors $W_{[1]}^\alpha$ and $W_{[L]}^\alpha$ at the two ends take the form
\begin{equation}\label{w1}
W_{[1]}^\alpha=
\begin{pmatrix}
F_{\alpha,1}^{\,\,M_\alpha}\; &
C_{M_\alpha}^{1}F_{\alpha,1}^{\,\,M_\alpha-1}\; &
C_{M_\alpha}^{2}F_{\alpha,1}^{\,\,M_\alpha-2}\; &
\cdots &
\mathds{1}
\end{pmatrix}\,,\nonumber\\
\end{equation}
and
\begin{equation}\label{wL}
W_{[L]}^\alpha=\begin{pmatrix}
\mathds{1} \\
F_{\alpha,L} \\
F_{\alpha,L}^{\,\,2} \\
\;\vdots \\
F_{\alpha,L}^{\,\,M_\alpha}
\end{pmatrix}\,,\nonumber\\
\end{equation}
with $C_{M_\alpha}^{k}$ being the binomial coefficient, and $\mathds{1}$ being the identity matrix,
and the bulk matrices
$W_{[j]}^\alpha$ at the lattice sites $j=2,\;\ldots, \;L-1$ take the form
\begin{equation}\label{wj}
W_{[j]}^\alpha=
\begin{pmatrix}
\,\mathds{1} & & & \vspace{5pt} \\
F_{\alpha,j}& \;\mathds{1} & & & \vspace{5pt} \\
F_{\alpha,j}^{\,\,2} & C_{2}^{1}F_{\alpha,j} & \;\mathds{1} & & \\
\;\vdots & & & \cdots & \\
F_{\alpha,j}^{\,\,M_\alpha} & C_{M_\alpha}^{1}F_{\alpha,j}^{\,\,M_\alpha-1} & C_{M_\alpha}^{2}F_{\alpha,j}^{\,\,M_\alpha-2} & \cdots & \mathds{1}
\end{pmatrix}\,.\nonumber\\
\end{equation}
Actually, $W_{[j]}^\alpha$ ($j=2$, \ldots, $L-1$) are either uniform or staggered, depending on the nature of the symmetry group $G$. We remark that such a MPO representation has already been exploited in a different guise~\cite{yuping}, without mentioning any connection with a lowering operator for any symmetry group.
Third,  an exact MPS representation for a scale-invariant state $|\psi(L,M_1,\ldots,M_r)\rangle_p=\prod_{\alpha=1}^r F_\alpha^{\,\,M_\alpha}|\psi_0\rangle_q$ simply follows from contracting the MPO representations for all the $r$ lowering operators with a factorized state that represents  the highest weight state or a generalized highest weight state (if any), as visualized in Fig.~\ref{mps}(iii). We stress that the period $p$ is not necessarily identical to the period $q$, if the symmetry group is staggered.
As a result, we are led to an exact MPS representation for this type of scale-invariant states, as pictorized in Fig.~\ref{mps}(iv). Such a representation  is efficient, in the sense that the bond dimension only scales as $L$.

We turn to a translation-invariant MPS representation for a scale-invariant state under PBCs.
Suppose an exact MPS product state representation under OBCs is known, one may turn it into an exact MPS representation under PBCs. We relegate the details of this construction to Sec. A of the Supplementary Material (SM). A consequence to be drawn from this construction is that the norm for an exact MPS representation under PBCs is identical to an exact MPS representation under OBCs. This offers an alternative way to understand why the entanglement entropy does not depend on what types of the boundary conditions are adopted for a scale-invariant state.

Our prescription leads to an exact MPS representation for a type of scale-invariant states arising from SSB with type-B GMs, which may be turned into a canonical form~\cite{1d2dTN}. This offers a powerful means to extract a variety of physical properties for a specific model under investigation. In particular, it is straightforward to evaluate the norms $Z(L,M_1,\ldots,M_r)$ and $Z_p(L,M_1,\ldots,M_r)$. Here, $Z(L,M_1,\ldots,M_r)$ and $Z_p(L,M_1,\ldots,M_r)$ represent the norms for scale-invariant states $|\psi(L,M_1,\ldots,M_r)\rangle$ and $|\psi(L,M_1,\ldots,M_r)\rangle_p$, depending on whether a degenerate ground state is generated from a generalized highest weight state, with the period being $p$. As a convention, we always keep the notation $Z(L,M_1,\ldots,M_r)$ to denote $Z_p(L,M_1,\ldots,M_r)$ if $p=1$. This in turn allows to perform a universal finite system-size scaling analysis of the entanglement entropy~\cite{finitesize}.

If the system is partitioned into a block $B$ and its environment $E$, with the block consisting of $n$ lattice sites, and the environment $E$ consisting of the other $L-n$ lattice sites, then one may introduce the reduced density matrix $\rho(L,n,M_1,\ldots,M_r)$ for the block  $B$. Generically,
the entanglement entropy $S_L(n,M_1,\ldots,M_r)$ for $|L,M_1,\ldots,M_r\rangle_p=1/Z_p(L,M_1,\ldots,M_r)|\psi(L,M_1,\ldots,M_r)\rangle_p$ takes the form~\cite{FMGM},
	\begin{align}
	S_L(n,M_1,\ldots,M_N)=& -\sum_{k_1,\ldots,k_r} \Lambda(L,k_1,\ldots,k_r,M_1,\ldots,M_r) \nonumber\\
	&\log_{2}\Lambda(L,k_1,\ldots,k_r,M_1,\ldots,M_r),
	\label{snk}
	\end{align}
where $\Lambda(L,k_1,\ldots,k_r,M_1,\ldots,M_r)$ denote the eigenvalues of the reduced density matrix $\rho(L,n,M_1,\ldots,M_r)$
	\begin{align*}
	&\Lambda(L,k_1,\ldots,k_r,M_1,\ldots,M_r)=\prod_{\alpha=1}^N(C_{M_\alpha}^{k_\alpha})^2 \times \nonumber \\
	&	\frac{[Z_p(n,k_1,\ldots,k_r)]^2[Z_p(L-n,M_1-k_1,\ldots,M_N-k_r)]^2}{[Z_p(L,M_1,\ldots,M_r)]^2}.	\nonumber\\
	\end{align*}
This observation is meaningful, given that it is a formidable task to evaluate the norms for this type of scale-invariant states in the representation-theoretic context~\cite{FMGM}. This in turn allows us to extract the number of type-B GMs $N_B$.
More precisely,  for a nonzero filling $f$, the entanglement entropy $S_{\!\!f}(L,n)$ for a block, with the block size being $n$, takes the form~\cite{finitesize}
\begin{equation}
	S_{\!\!f} (L,n)=\frac{N_B}{2} \log_2\frac{n(L-n)}{L} +S_{\!\!f0}.
	\label{slnf}
\end{equation}
Here, $f$ is defined as $f \equiv (f_1,f_2,\ldots, f_r)$, with $f_{\alpha} =M_{\alpha}/L$ ($\alpha = 1,2,\ldots,r)$, and $S_{\!\!f0}$ is an additive nonuniversal constant. Note that
the prefactor is just half the number of type-B GMs $N_B$. We stress that this finite system-size scaling relation reproduces the logarithmic scaling relation with the block size $n$ in the thermodynamic limit.

\section{Three illustrative examples}

Our prescription for constructing an exact MPS representation is explicitly carried out for highly degenerate ground states arising from SSB with type-B GMs in the ${\rm SU}(2)$ spin-$s$ Heisenberg ferromagnetic model, the ${\rm SU}(2s+1)$ ferromagnetic model, and the staggered ${\rm SU}(3)$ spin-1 ferromagnetic biquadratic model.
Here,  we have restricted ourselves to the final MPS representations. The details of a MPS representation for the highest weight state and a MPO representation for a power of the lowering operator(s) may be found for each of the three models in Sec. B and Sec. C of the SM.

\subsection{The ${\rm SU}(2)$ spin-$s$ ferromagnetic Heisenberg model}

The ${\rm SU}(2)$ spin-$s$ ferromagnetic Heisenberg model under PBCs is described by the Hamiltonian
\begin{equation}
\mathscr{H}=-\sum_{j=1}^{L}\textbf{S}_j\cdot \textbf{S}_{j+1}, \label{su2ham}
\end{equation}
where $\textbf{S}_j=(S^x_j,S^y_j,S^z_j)$, with $S^x_j$, $S^y_j$, $S^z_j$ being the spin-$s$ operators at the $j$-th lattice site.
The spin-$s$ Heisenberg model is exactly solvable, as far as its ground state subspace is concerned. However, it becomes exactly solvable by means of the Bethe ansatz for $s=1/2$.
The highly degenerate ground states arise from SSB: ${\rm SU}(2) \rightarrow {\rm U}(1)$, so there is one type-B GM: $N_B = 1$.

Following the prescription,  the highly degenerate ground states $|L,M\rangle=1/Z(L,M)(S_-)^{M}|\psi_0\rangle$, with the highest weight state $|\psi_0\rangle=\otimes_{l=1}^L\{s\}_l\rangle$,  admit an exact MPS representation,
\begin{equation}
|L,M\rangle = \sum_{s_1,\ldots,s_j,\ldots,s_L}  (B_{1}^{s_1}\cdots B^{s_j}\cdots B_{L}^{s_L})|s_1\cdots s_j\cdots s_L\rangle,
\label{su2spinslm}
\end{equation}
where the explicit expressions of the two vectors $B_{1}^{s_1}$ and $B_{L}^{s_L}$ at the two ends as well as
and the matrices $B^{s_j}$ at the lattice sites $j=2,\;\ldots, \;L-1$, which are identical, may be found in Sec.\;D of the SM.

\subsection{The ${\rm SU}(2s+1)$ ferromagnetic model}

The ${\rm SU}(2s+1)$ ferromagnetic model under PBCs is described by the Hamiltonian
\begin{equation}
\mathscr{H}=-\sum_{j}^L P_{j\;j+1}, \label{HsuNp1}
\end{equation}
where $P$ is the permutation operator, which may be realized in terms of the spin-$s$ operators $\textbf{S}_j=(S^x_j,S^y_j,S^z_j)$, with $S^x_j$, $S^y_j$, $S^z_j$ being the spin-$s$ operators at the $j$-th lattice site.
We remark that the ${\rm SU}(2s+1)$ ferromagnetic model is exactly solvable by means of the nested Bethe ansatz~\cite{sutherland}.
In particular, it becomes one exactly solved point for the bilinear-biquadratic model~\cite{TB,daibb}.
The highly degenerate ground states $|L,M_1,\ldots,M_{2s}\rangle$ arise from SSB: ${\rm SU}(2s+1) \rightarrow {\rm SU}(2s) \times {\rm U}(1)$ successively. Hence, we have
$2s$ type-B GMs: $N_B = 2s$.

Following the prescription, the highly degenerate ground states
$	|L,M_1,\ldots,M_{2s}\rangle
	={1}/{Z(L,M_1,\ldots,M_{2s})}\prod_{\alpha=1}^{2s}F_\alpha^{\,\,M_\alpha}|\psi_0\rangle$,
with the highest weight state $|\psi_0\rangle=\otimes_{l=1}^L\{s\}_l\rangle$, admit an exact MPS representation,
\begin{equation}
|L,M_1,\ldots,M_{2s}\rangle = \sum_{s_1,\ldots,s_j,\ldots,s_L}  (B_1^{s_1}\cdots B^{s_j}\cdots B_L^{s_L})|s_1\cdots s_j\cdots s_L\rangle,
\label{su2sp1lm}
\end{equation}
where the two vectors $B_1^{s_1}$ and $B_L^{s_L}$ at the two ends
and the matrices $B^{s_j}$ at the lattice sites $j=2,\;\ldots, \;L-1$, which are identical, may be found in Sec.\;D of the SM.

\subsection{The staggered ${\rm SU}(3)$ spin-1 ferromagnetic biquadratic model}

The staggered ${\rm SU}(3)$ spin-1 ferromagnetic biquadratic model under PBCs is described by the Hamiltonian
\begin{equation}
\mathscr{H}=\sum_{j}^L{\left(\textbf{S}_j \cdot \textbf{S}_{j+1}\right)^2}, \label{hambq}
\end{equation}
where $\textbf{S}_j=(S^x_j,S^y_j,S^z_j)$, with $S^x_j$, $S^y_j$, $S^z_j$ being the spin-$1$ operators at the $j$-th lattice site.
The model (\ref{hambq}) is exactly solvable~\cite{barber,klumper}. Indeed, it constitutes (up to an additive constant) a representation of the Temperley-Lieb algebra~\cite{tla,baxterbook,martin}, and thus follows from a solution to the Yang-Baxter equation~\cite{baxterbook,sutherlandb,mccoy}.
Note that it is peculiar, in the sense that the ground states are highly degenerate, exponential with the system size $L$, thus leading to non-zero residual entropy~\cite{spins},
in sharp contrast to the ${\rm SU}(2)$ spin-$s$ ferromagnetic model (\ref{su2ham}) and the ${\rm SU}(2s+1)$ ferromagnetic model (\ref{HsuNp1}).
Remarkably, the highly degenerate ground states arise from SSB: ${\rm SU}(3) \rightarrow {\rm U}(1) \times {\rm U}(1)$, with the number of type-B GMs being two: $N_B=2$, and the ground state degeneracies constitute two Fibonacci-Lucas sequences under OBCs and PBCs~\cite{golden} (also cf.~\cite{spins,saleur}).

Following the prescription,  the highly degenerate ground states $|L,M_1,M_2\rangle_2=1/Z_2(L,M_1,M_2)F_1^{M_1}F_2^{M_2}|\psi_0\rangle$, with the highest weight state $|\psi_0\rangle=\otimes_{l=1}^L\{1\}_l\rangle$, admit an exact MPS representation,
\begin{align}
	|L,M_1,M_2\rangle_2 = &\sum_{s_1,\ldots,s_L}  (B_1^{s_1}\cdots B_{[1]}^{s_{2j-1}}B_{[2]}^{s_{2j}}\cdots B_L^{s_L})\nonumber \\
	&|s_1\cdots s_{2j-1}s_{2j}\cdots s_L\rangle,
	\label{lm1m2}
\end{align}
where the two vectors $B_{1}^{s_1}$ and $B_{L}^{s_L}$ at the two ends
and the matrices $B_{[1]}^{s_{2j-1}}$ at the $(2j-1)$-th lattice sites, with $j=2,\;\ldots, \;L/2$, may be found in Sec.\;D of the SM.

Following the prescription,  the highly degenerate ground states $|L,M_2,M_3\rangle_4=1/Z_4(L,M_2,M_3)F_2^{M_2}F_3^{M_3}|\psi_0\rangle_4$, with a generalized highest weight state $|\psi_0\rangle_4=|\otimes_{l=1}^{L/4}\{1110\}_l\rangle$,  admit an exact MPS representation,
\begin{align}
	&|L,M_2,M_3\rangle_4 = \sum_{s_1,\ldots,s_L} (B_{1}^{s_1}B_{[2]}^{s_2}B_{[3]}^{s_3}B_{[4]}^{s_4}\cdots \nonumber \\
	& B_{[1]}^{s_{4m-3}}B_{[2]}^{s_{4m-2}}B_{[3]}^{s_{4m-1}}B_{[4]}^{s_{4m}} \cdots B_{[1]}^{s_{L-3}}B_{[2]}^{s_{L-2}}B_{[3]}^{s_{L-1}}B_L^{s_L}) \nonumber \\
	&|s_1s_2s_3s_4\cdots s_{4m-3}s_{4m-2}s_{4m-1}s_{4j}\cdots s_{L-3}s_{L-2}s_{L-1}s_L\rangle,
	\label{Lm2m3g}
\end{align}
where the two vectors $B_1^{s_1}$ and $B_L^{s_L}$ at the two ends
and the matrices $B_{[1]}^{s_{4m-3}}$ at the $(4m-3)$-th lattice sites, with $m=2,3,\;\ldots, \;L/4$,
the matrices $B_{[2]}^{s_{4m-2}}$ at the $(4m-2)-th$ lattice sites, with $m=1,2,\;\ldots, \;L/4$,
the matrices $B_{[3]}^{s_{4m-1}}$ at the $(4m-1$)-th lattice sites, with  $m=1,2,\;\ldots, \;L/4$,
and the matrices $B_0^{s_{4m}}$ at the $4m$-th lattice sites, with $m=1,2,\;\ldots, \;L/4-1$, may be found in Sec.\;D of the SM.

\section{Universal finite system-size scaling for the entanglement entropy}

We perform a universal finite system-size scaling analysis for the entanglement entropy to extract the number of type-B GMs $N_B$, according to Eq.(\ref{slnf}).
Here, both $N_B$ and $S_{\!\!f0}$ are treated as the fitting parameters when
the best linear fit is carried out for the three illustrative models.

\begin{figure}[htb]
	\centering
	\includegraphics[width=0.5\textwidth]{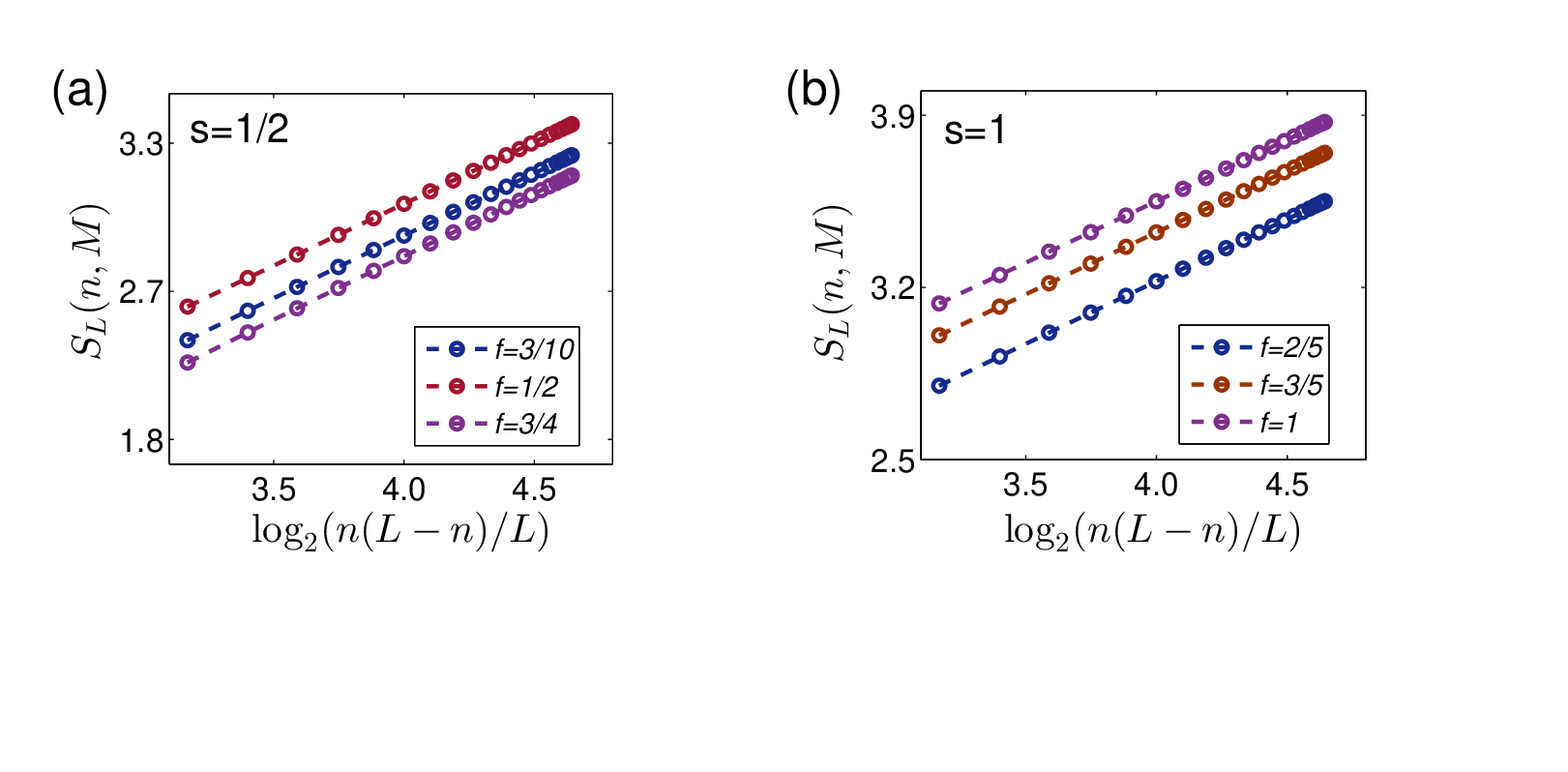}
	\caption{ The entanglement entropy $S_L(n,M)$ vs $\log_2 [n(L-n)/L]$ for the highly degenerate ground states  $|L,M\rangle$ in the ${\rm SU}(2)$
spin-$s$ ferromagnetic Heisenberg model, with the system size $L=100$,  when $n$ varies from 10 to 50.
(a) For $s=1/2$, the filling $f$ is chosen to be $f=3/10$, $1/2$ and $3/4$.
(b) For $s=1$, the filling $f$ is chosen to be $f=2/5$, $3/5$ and $1$.}
	\label{comparesu2spins}
\end{figure}

In Fig.\;\ref{comparesu2spins}, we plot the entanglement entropy $S_L(L,M)$ vs $\log_2 [n(L-n)/L]$ for the highly degenerate ground states $|L,M\rangle$, with the filling $f=M/L$,
when $L=100$ and $n$ ranges from 10 to 50.  The best linear fit yields that the number of type-B GMs $N_B$ is one,
with the relative errors, as measured by a deviation of $N_B$ from
the exact result $N_B=1$,  are less than $2.5\%$ (cf. Table~\ref{tab1} in Sec. E of the SM).

\begin{figure}[htb]
	\centering
	\includegraphics[width=0.45\textwidth]{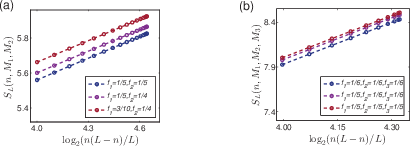}
	\caption{	(a) The entanglement entropy $S_L(n,M_1,M_2)$ vs $\log_2 [n(L-n)/L]$ for the highly degenerate ground states  $|L,M_1,M_2\rangle$ in the ${\rm SU}(3)$ ferromagnetic model, with the system size $L=100$,  when $n$ varies from 10 to 50. The fillings are chosen to be $f_1=1/5$ and $f_2=1/5$, $f_1=1/5$ and $f_2=1/4$, and $f_1=3/10$ and $f_2=1/4$, respectively.
	(b) The entanglement entropy $S_L(n,M_1,M_2,M_3)$ vs  $\log_2 [n(L-n)/L]$ for the highly degenerate ground states $|L,M_1,M_2, M_3\rangle$ in the ${\rm SU}(4)$ ferromagnetic model, with the system size $L=80$,  when $n$ varies from 20 to 40.  The fillings are chosen to be $f_1=1/6$, $f_2=1/6$ and $f_3=1/6$, $f_1=1/6$, $f_2=1/5$ and $f_3=1/5$, and $f_1=1/5$, $f_2=1/5$ and $f_3=1/5$, respectively.
	}
	\label{comparesu3su4}
\end{figure}

In Fig.\;\ref{comparesu3su4}(a), we plot the entanglement entropy $S_L(n,M_1,M_2)$ vs $\log_2 [n(L-n)/L]$ for $|L,M_1,M_2\rangle$, with the fillings $f_1=M_1/L$ and $f_2=M_2/L$
being $f_1=1/5$ and $f_2=1/5$, $f_1=1/5$ and $f_2=1/4$, and $f_1=3/10$ and $f_2=1/4$, respectively, when $L=100$ and $n$ ranges from 10 to 50.
The best linear fit yields that the number of type-B GMs $N_B$ is two,
with the relative errors, as measured by a deviation of $N_B$ from
the exact result $N_B=2$,  are less than $3\%$ (cf. Table~\ref{tab2} in Sec. E of the SM).

In Fig.\;\ref{comparesu3su4}(b), we plot the entanglement entropy $S_L(n,M_1,M_2,M_3)$ vs $\log_2 [n(L-n)/L]$ for $|L,M_1,M_2, M_3\rangle$, with the filling factors $f_1=M_1/L$, $f_2=M_2/L$ and $f_{3}=M_3/L$ being $f_1=1/6$, $f_2=1/6$ and $f_3=1/6$, $f_1=1/6$, $f_2=1/5$ and $f_3=1/5$, and $f_1=1/5$, $f_2=1/5$ and $f_3=1/5$, respectively,
when $L=80$ and $n$ ranges from 20 to 40. The best linear fit yields that the number of type-B GMs $N_B$ is three,
with the relative errors, as measured by a deviation of $N_B$ from
the exact result $N_B=3$,  are less than $4\%$ (cf. Table~\ref{tab2} in Sec. E of the SM).

\begin{figure}[htb]
	\centering
	\includegraphics[width=0.45\textwidth]{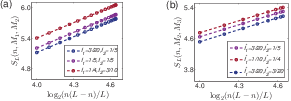}
	\caption{ The entanglement entropy $S_L(n,M_1,M_2)$ vs $\log_2 [n(L-n)/L]$ for the highly degenerate ground states $|L,M_1,M_2\rangle_2$ in the staggered ${\rm SU}(3)$ spin-1 ferromagnetic biquadratic model, when the fillings $f_1$ and $f_2$ are chosen to be $f_1=3/20$ and $f_2=1/5$, $f_1=1/5$ and $f_2=1/5$, and $f_1=1/4$ and $f_2=3/10$, respectively.
	(b) The entanglement entropy $S_L(n,M_2,M_3)$ vs $\log_2 [n(L-n)/L]$ for the highly degenerate ground states $|L,M_1^*,M_2^*\rangle_4$ in the staggered ${\rm SU}(3)$  spin-1 ferromagnetic biquadratic model, when the fillings $f_1^*$ and $f_2^*$ are chosen to be  $f_1^*=3/20$ and $f_2^*=3/20$, $f_1^*=3/20$ and $f_2^*=1/5$ and $f_1^*=1/10$ and $f_2^*=1/4$, respectively.
	In both cases, $L=100$ and $n$ ranges from 20 to 50.}
	\label{comparestsu3}
\end{figure}

In Fig.~\ref{comparestsu3}, we plot the entanglement entropy $S_L(n,M_1,M_2)$ and the entanglement entropy $S_L(n,M_2,M_3)$ vs  $\log_2 [n(L-n)/L]$ for  $|L,M_1,M_2\rangle_2$ and  $|L,M_1^*,M_2^*\rangle_4$, respectively, where the fillings $f_1=M_1/L$ and $f_2=M_2/L$, and the fillings $f_1^*=M_1^*/L$ and $f_2^*=M_2^*/L$, with $M_1^*=M_2$ and $M_2^*=M_3$,
when $L=100$ and $n$ ranges from 20 to 50.
Here, we select the fillings (a) $f_1=3/20$ and $f_2=1/5$, $f_1=1/5$ and $f_2=1/5$, and $f_1=1/4$ and $f_2=3/10$ and (b)  $f_1^*=3/20$ and $f_2^*=3/20$, $f_1^*=3/20$ and $f_2^*=1/5$ and $f_1^*=1/10$ and $f_2^*=1/4$, respectively.
The best linear fit yields that the number of type-B GMs $N_B$ is two,
with the relative errors, as measured by a deviation of $N_B$ from
the exact result $N_B=2$,  are less than $4\%$ (cf. Table~\ref{tab3} in Sec. E of the SM).

%

\section{Summary}

An exact MPS representation has been constructed for scale-invariant states, which appear as highly degenerate ground states arising from SSB with type-B GMs in one-dimensional quantum many-body systems. Such a representation offers a powerful means for evaluating the norms of highly degenerate ground states. This in turn allows us to perform
a universal finite system-size scaling analysis of the entanglement entropy for one-dimensional quantum many-body systems undergoing SSB with type-B GMs. Moreover, our generic scheme allows us to turn a MPS representation for a scale-invariant state under OBCs into  that for a scale-invariant state under PBCs, thus offering a vivid explanation for an observation that
the entanglement entropy does not depend on what types of the boundary conditions are adopted.

\section{Acknowledgements}
We are grateful to Murray Batchelor and John Fjaerestad  for helpful discussions.
I.P.M. acknowledges funding from the National Science and Technology Council (NSTC) Grant No. 122-2811-M-007-044.

\newpage
\onecolumngrid
\newpage
\section*{Supplementary Material}
\twocolumngrid
\setcounter{page}{1}
\setcounter{equation}{0}
\setcounter{figure}{0}
\renewcommand{\theequation}{S\arabic{equation}}
\renewcommand{\thefigure}{S\arabic{figure}}
\renewcommand{\bibnumfmt}[1]{[S#1]}
\renewcommand{\citenumfont}[1]{S#1}

\subsection{A: Turning a MPS representation under OBCs into a  translation-invariant MPS representation under PBCs}

Once a MPS representation for highly degenerate ground states is known under OBCs, it is natural to turn it into a  translation-invariant MPS representation under PBCs, given the models under investigation are translation-invariant.   In fact, the Hamiltonian $\mathscr{H}$
commutes with the one-site translation operation $\tau$ under PBCs. In practice, it is important to take full advantage of a translation-invariant MPS representation.
Here, we take the spin-$s$ Heisenberg ferromagnetic states $|L,M\rangle$ in Eq.~(\ref{su2spinslm}) as an example.
However, its extension to a translation-invariant MPS representation for ground states in other models under PBCs, even if the one-site translation operation $\tau$ is spontaneously broken, is straightforward.
\begin{figure}[htpt]
	\includegraphics[angle=0,totalheight=5cm]{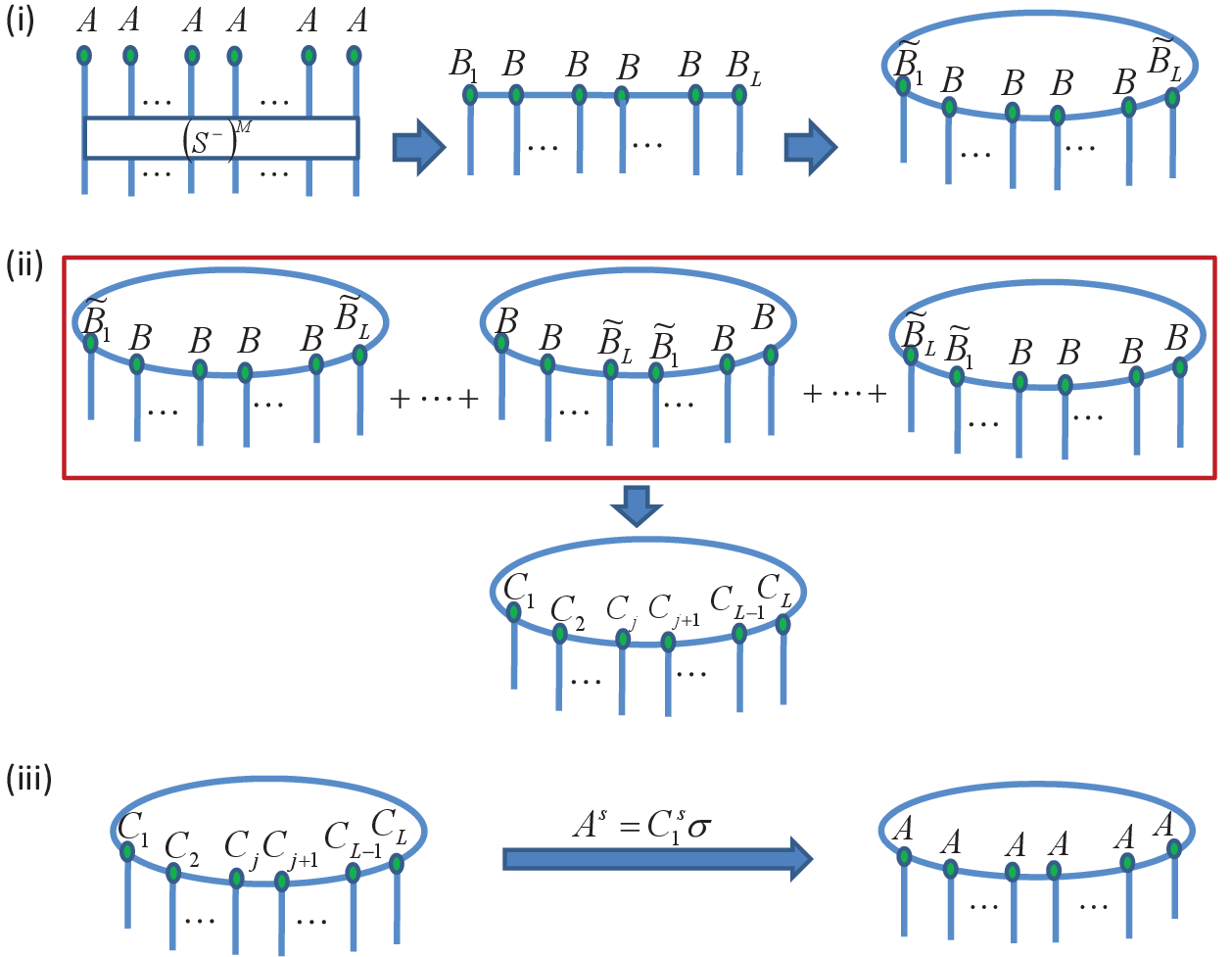}
	\caption{ A translation-invariant MPS representation for $|L,M\rangle$ under PBCs. (i) A MPS representation for $|L,M\rangle$ under OBCs, generated from the action of a MPO representation on a factorized state, is converted to a MPS representation for $|L,M\rangle$ under PBCs, with an additional bond connecting the lattice site 1
		and the lattice site $L$. Here, $\tilde{B}_{1}$ and $\tilde{B}_{L}$ are  constructed from two vectors $B_{1}$ and $B_{L}$ by adding zeros to turn them into two matrices.
		We stress that $|L,M\rangle$ is permutation-invariant, thus implying that
		$|L,M\rangle$ is invariant under  a $Z_L$ cyclic symmetry group, generated by the permutation $\sigma$ that permutes $(1,2, \ldots, L)$ to   $(2, \ldots, L,1)$,
		satisfying $\sigma^L=I$, with $I$ being the identity matrix. (ii) The translation-invariant MPS representation for $|L,M\rangle$ under PBCs is constructed as a sum of $L$ copies of the MPS representations in (i), which may be merged into one single MPS representation, if the bond dimension $\chi$ is increased to $L\chi$, with $\chi=M+1$. (iii)  Take advantage of a representation of $Z_L$ in a vector space, consisting of all the block-diagonal $L\chi \times L\chi$ matrices, with  the number of diagonal blocks being $L$ and each block being a $\chi \times \chi$ matrix, we are able to turn  a MPS representation for $|L,M\rangle$ under OBCs to a translation-invariant MPS for $|L,M\rangle$ under PBCs.
	}\label{pbcmpslm}
\end{figure}

Specifically, a translation-invariant MPS representation for a degenerate ground state $|\psi\rangle$, arising from SSB with type-B GMs, in a quantum many-body system under PBCs, is constructed as follows.

First, a MPS representation for $|L,M\rangle$ under OBCs, generated from the action of a MPO representation on a factorized state, is converted to a MPS representation for $|L,M\rangle$ under PBCs, with an additional bond connecting the lattice site 1
and the lattice site $L$, as shown in Fig.~\ref{pbcmpslm}(i). Here, $\tilde{B}_{1}$ and $\tilde{B}_{L}$ are  constructed from two vectors $B_{1}$ and $B_{L}$ by adding zeros to turn them into two matrices.  That is, $\tilde{B}_{1}$ is a $\chi\times\chi$ matrix, with the first row being identical to $B_1$, and all the other rows being zero vectors, with $\chi=M+1$.
Similarly, $\tilde{B}_{L}$ is a $\chi\times\chi$ matrix,   with the first column being identical to $B_L$, and all the other columns being zero vectors. Therefore, it is legitimate to take a trace operation:
\begin{equation}
	|L,M\rangle =\sum_{s_1,\ldots,s_j,\ldots,s_L} {\rm Tr} (\tilde{B}_1^{s_1}\cdots B^{s_j}\cdots\tilde{B}_L^{s_L})|s_1\cdots s_j\cdots s_L\rangle.
\end{equation}
However, this MPS representation is not translation-invariant. Hence, an extra step needs to be taken to recover the translational invariance.

Second,  the translation-invariant MPS representation for $|L,M\rangle$ under PBCs is constructed as a sum of $L$ copies of the non-translation-invariant MPS representation, as shown in Fig.~\ref{pbcmpslm}(ii). Indeed, it may be merged into one single MPS representation, if the bond dimension $\chi$ is increased to $L\chi$. Mathematically, we have
\begin{align}
	|\psi\rangle &=\sum_{s_1,\ldots,s_j,\;\ldots,s_L}  {\rm Tr} (\tilde{B}_1^{s_1}\cdots B^{s_j}\cdots\tilde{B}_L^{s_L})|s_1\cdots s_j\cdots s_L\rangle+\cdots  \nonumber \\
	&+\sum_{s_1,\ldots,s_j,\;\ldots,s_L}  {\rm Tr} (\tilde{B}_1^{s_1}\cdots B^{s_j}\cdots\tilde{B}_L^{s_L})|s_j\cdots s_Ls_1\cdots s_{j-1}\rangle+\cdots \nonumber \\
	&+\sum_{s_1,\ldots,s_j,\;\ldots,s_L}  {\rm Tr} (\tilde{B}_1^{s_1}\cdots B^{s_j}\cdots\tilde{B}_L^{s_L})|s_Ls_1\cdots s_j\cdots s_{L-1}\rangle,\nonumber \\
	&=\sum_{s_1,\ldots,s_j,\;\ldots,s_L} {\rm Tr} (C_1^{s_1}
	\cdots C_j^{s_j}\cdots C_L^{s_L})|s_1\cdots s_j\cdots s_L\rangle,
\end{align}
Here, $C_1^{s_1}$, $C_j^{s_j}$ ($j=2$, \ldots, $L-1$) and $C_L^{s_L}$ take the form
\begin{equation*}
	C_{1}^{s_1} = \begin{pmatrix}\label{c1}
		\tilde{B}_{1}^{s_1} & \cdots & O&\cdots &O\\
		\vdots & \vdots & \vdots&\cdots &\vdots\\
		O &\cdots & B^{s_1} &\cdots  & O\\
		\vdots & \vdots &\vdots& \cdots &\vdots\\
		O &  \cdots & O &\cdots& \tilde{B}_{L}^{s_1} \\
	\end{pmatrix},
\end{equation*}
\begin{equation*}
	C_j^{s_j} = \begin{pmatrix}\label{cj}
		B^{s_{j}} & \cdots & O & O& \cdots & O\\
		\vdots & \vdots & \vdots & \vdots &\cdots &\vdots\\
		O & \cdots & \tilde{B}_{L}^{s_j} &O&\cdots &O\\
		O & \cdots & O & \tilde{B}_{1}^{s_j} &\cdots  & O\\
		\vdots & \vdots & \vdots & \vdots  &\cdots & \vdots\\
		O &  \cdots & O & O &\cdots & B^{s_j} \\
	\end{pmatrix},
\end{equation*}
and
\begin{equation*}
	C_L^{s_L} = \begin{pmatrix}\label{cl}
		\tilde{B}_L^{s_L} & O & \cdots  & O\\
		O &  \tilde{B}_1^{s_L} &\vdots  &\vdots\\
		\vdots& \vdots & \vdots & \vdots\\
		O & \cdots & \cdots & B^{s_L}\\
	\end{pmatrix},
\end{equation*}
where $O$ is a $\chi L\times\chi L$ matrix, with all the entries being zero.

Third, one may take advantage of a representation of $Z_L$ in a vector space, consisting of all the block-diagonal $L\chi \times L\chi$ matrices, with  the number of diagonal blocks being $L$ and each block being a $\chi \times \chi$ matrix, in order to turn  a MPS representation for $|L,M\rangle$ under OBCs to a translation-invariant MPS for $|L,M\rangle$ under PBCs, as shown in Fig.~\ref{pbcmpslm}(iii). Assume that $\sigma$ is the generator of the $Z_L$ cyclic symmetry group, satisfying $\sigma^L=	\mathds{1}$, with $\mathds{1}$ being the identity matrix, then a representation of $Z_L$ in such a vector space is realized in terms of $\mathbf{\sigma}$ in the form,
\begin{equation}
	\mathbf{\sigma} = \begin{pmatrix}\label{equation: sigma}
		O & \mathds{1} & O &\cdots  & O\\
		O & O &	\mathds{1} &\cdots & O\\
		\vdots & \vdots & \vdots &\vdots &O \\
		O & O & O & \cdots & 	\mathds{1}\\
		\mathds{1} & O & O&\cdots& O \\
	\end{pmatrix},
\end{equation}
We stress that $C_j^s$ are connected via $\sigma$
\begin{equation}
	C_j^s=\sigma^{j-1}C_1^{s}\sigma^{-j+1}. \nonumber\\
\end{equation}
As a result, we have
\begin{align*}
	|\psi\rangle = &\sum_{s_1,s_2, \ldots,s_j, \ldots,s_L} {\rm Tr} (C_1^{s_1}\sigma C_1^{s_2}\sigma
	\cdots \sigma C_1^{s_j}\sigma\cdots \sigma C_1^{s_L}\sigma)\nonumber \\
	&|s_1s_2\cdots s_j\cdots s_L\rangle.
\end{align*}
The simplest choice is to set $A^{s}=C_1^s\sigma$.
That is, we are able to re-express $|L,M\rangle$ in terms of $A^s$ as follows
\begin{equation}
	|L,M\rangle \propto |\psi\rangle \!= \!\sum_{s_1,\ldots,s_j,\ldots,s_L} {\rm Tr} (A^{s_1}\cdots A^{s_j}\cdots A^{s_L})|s_1\cdots s_j\cdots s_L\rangle.
\end{equation}
Hence, one is able to turn a MPS representation for $|L,M\rangle$ under OBCs into that under PBCs for any  $L$ and $M$. Our construction above explains why the norm for $|L,M\rangle$
remains to be the same, when OBCs are varied to PBCs.

This construction works for highly degenerate ground states in one-dimensional quantum many-body systems undergoing SSB with type-B GMs. Physically, this is due to the fact that highly degenerate ground states under OPCs are invariant under an {\it emergent} cyclic permutation symmetry operation $P_{12}P_{23}\cdots,P_{L-1,L}$, where
$P_{kk+1}$ ($k=1,\cdots,L-1$) denote the generators of the permutation group $S_L$, if the one-site translation operation $\tau$ is not spontaneously broken. In fact, the action of this emergent cyclic permutation symmetry operation on a degenerate ground state under OPCs is identical to the action of the one-site translation operation $\tau$ on the degenerate ground state under PBCs.
If the one-site translation operation $\tau$ is spontaneously broken, we resort to a MPS representation involving a unit cell with the size being $p$. Accordingly, it is translation-invariant under $\tau^p$. Meanwhile, an emergent cyclic permutation symmetry group $S_{L/p}$ emerges for a degenerate ground state under OBCs, consisting of all the permutation operations with respect to the unit cells.

\subsection{B: The highest weight state and generalized highest weight states: matrix product state representations}

According to our prescription, the first step is to turn the highest weight state (and generalized highest weight states, if any) into a trivial matrix product state representation.

\subsubsection{1. The ${\rm SU}(2)$ spin-$s$ ferromagnetic Heisenberg model}

For the ${\rm SU}(2)$ spin-$s$ ferromagnetic Heisenberg model, the highly degenerate ground states $|L,M\rangle$ are generated from the repeated action of the lowering operator $S^-$ on the highest weight state $|\psi_0 \rangle$, where $S^- = \sum _j  S^-_j$, with $ S^-_j=(S_{x,j}- iS_{y,j})/\sqrt{2}$:
\begin{equation}
	|L,M\rangle=\frac{1}{Z(L,M)}(S^-)^{\,\,M}|\psi_0 \rangle.\label{lmspins}
\end{equation}
Here, $Z(L,M)$ is a normalization factor, and the highest weight state $|\psi_0 \rangle =|\otimes_{l=1}^L\{s\}_l\rangle$ is a factorized state. Hence, its MPS representation takes the form
\begin{equation}
	|\psi_0 \rangle= \sum_{s_1,\ldots,s_j,\ldots,s_L}  (A^{s_1}\cdots A^{s_j}\cdots A^{s_L})|s_1\cdots s_j\cdots s_L\rangle.
		\label{hwsmps1}
\end{equation}
Here, $A$ is a $2s+1$-dimensional vector, if one takes the physical index into account. It may be specified as $A^{s_j}=1$ if $s_j=s$, and $A^{s_j}=0$ if $s_j<s$.
Meanwhile,  it is necessary to turn the power of the lowering operator $S^-$, i.e., $(S^-)^M$, into a MPO representation.

\subsubsection{2: The ${\rm SU}(2s+1)$ ferromagnetic model}

For the ${\rm SU}(2s+1)$ ferromagnetic model, the highly degenerate ground states $|L,M_1,\ldots,M_{2s}\rangle$ are generated from the repeated action of the lowering operators $F_{\alpha}$ on the highest weight state $|\psi_0\rangle=|\otimes_{l=1}^L\{1\}_l\rangle$:
\begin{equation}
	|L,M_1,\ldots,M_{2s}\rangle
	=\frac{1}{Z(L,M_1,\ldots,M_{2s})}\prod_{\alpha=1}^{2s}F_\alpha^{\,\,M_\alpha}|\otimes_{l=1}^L\{1\}_l\rangle.
	\label{gslmn}
\end{equation}
The highest weight state $|\psi_0 \rangle =|\otimes_{l=1}^L\{1\}_l\rangle$ is a factorized state. Hence, its MPS representation takes the form
\begin{equation}
	|\psi_0 \rangle  = \sum_{s_1,\ldots,s_j,\ldots,s_L} (A^{s_1}\ldots A^{s_j}\ldots A^{s_L})|s_1\ldots s_j\ldots s_L\rangle.
	\label{hwsmps2}
\end{equation}
Here, $A$ is a $2s+1$-dimensional vector, if one takes the physical index into account. It may be specified as $A^{s_j}=1$ if $s_j=s$, and $A^{s_j}=0$ if $s_j<s$.
Meanwhile,  it is necessary to turn the power of the lowering operator $F_\alpha^{\,\,M_\alpha}$ into a MPO representation.

\subsubsection{3: The staggered ${\rm SU}(3)$ spin-1 ferromagnetic biquadratic model}

For the staggered ${\rm SU}(3)$ spin-1 ferromagnetic biquadratic model, the highly degenerate ground states  $|L,M_1,M_2\rangle=1/Z_2(L,M_1,M_2)F_1^{M_1}F_2^{M_2}|\otimes_{l=1}^L\{1\}_l\rangle$ and $|L,M_2,M_3\rangle=1/Z_4(L,M_2,M_3)F_2^{M_2}F_3^{M_3}|\otimes_{l=1}^{L/4}\{1110\}_l\rangle$ are generated from the repeated action of the lowering operators $F_{\alpha}$ on the highest weight state $|\psi_0 \rangle =|\otimes_{l=1}^L\{1\}_l\rangle$ and  the generalized highest weight state $|\psi_0 \rangle =|\otimes_{l=1}^{L/4}\{1110\}_l\rangle$, respectively.

The highest weight state $|\psi_0 \rangle =|\otimes_{l=1}^L\{1\}_l\rangle$ is a factorized state.  Hence, its MPS representation takes the form
\begin{equation}
	|\psi_0 \rangle = \sum_{s_1,\ldots,s_j,\ldots,s_L}  (A^{s_1}\cdots A^{s_j}\cdots A^{s_L})|s_1\cdots s_j\cdots s_L\rangle.
		\label{hwsmps3}
\end{equation}
Here, $A$ is a $3$-dimensional vector, if one takes the physical index into account, specified as $A^{1}=1$ and $A^{s_j}=0$ when $s_j<1$.

The generalized highest weight state $|\psi_0 \rangle =|\otimes_{l=1}^{L/4}\{1110\}_l\rangle$, with the period $p$ being four, is a factorized state.  Hence, its MPS representation takes the form
\begin{equation}
	|\psi_0 \rangle = \sum_{s_1,\ldots,s_j,\ldots,s_L}  (A^{s_1}_1\cdots A^{s_j}_j\cdots A^{s_L}_L)|s_1\cdots s_j\cdots s_L\rangle.
		\label{hwsmps4}
\end{equation}
Here, $A^{s_j}_j$, with $j=4m-3,4m-2,4m-1$ ($m\geq 1$), are equal to $A$, with $A^{s}=1$ if $s=1$ and $A^{s}=0$ if $s<1$, whereas $A^{s_{4m}}_{4m}$ ($m\geq 1$) are equal to $A_0$, with $A^s_0=1$ if $s=0$ and $A^s_0=0$ if $s\neq0$.
Meanwhile,  it is necessary to turn the power of the lowering operator $F_\alpha^{\,\,M_\alpha}$ into a MPO representation.

\subsection{C: A matrix product operator representation for the power of a lowering operator}

To construct a MPS representation for highly degenerate ground states, a crucial step is to turn the  power of a lowering operator into a MPO representation.  This will be explicitly carried out for the three illustrative models.

\subsubsection{1. The ${\rm SU}(2)$ spin-$s$ Heisenberg ferromagnetic model}

For the ${\rm SU}(2)$ spin-$s$ Heisenberg ferromagnetic model, a MPO  representation for $(S^-)^M$ is  adapted from  that for a generic operator, developed in Ref.\;\cite{smyuping}, to the lowering operator $S^-$ of ${\rm SU}(2)$. Hence, we have

\begin{equation}
(S^-)^M=W_{[1]}\cdots W\cdots W_{[L]} \,,
\label{lo1}
\end{equation}
where the two matrices $W_1$ and $W_L$ at the two ends are
\begin{equation*}
W_{[1]}=
\begin{pmatrix}
	(S^-_j)^M\; &
	C_{M}^{1}(S^-_j)^{M-1}\; &
	C_{M}^{2}(S^-_j)^{M-2}\; &
	\cdots &
	\mathds{1}
\end{pmatrix}\,,
\end{equation*}
and
\begin{equation*}
W_{[L]}=\begin{pmatrix}
	\mathds{1} \\
	S^-_j \\
	(S^-_j)^2 \\
	\;\vdots \\
	(S^-_j)^M
\end{pmatrix}\,,
\end{equation*}
respectively, and the bulk matrices
at the lattice sites $j=2,\;\ldots,\; L-1$ are identical, denoted as $W$, which takes the form
\begin{equation*}
W=
\begin{pmatrix*}[l]
	\,\mathds{1} & & & \vspace{5pt} \\
	S^-_j & \;\mathds{1} & & & \vspace{5pt} \\
	(S^-_j)^2 & C_{2}^{1}S^-_j& \;\mathds{1} & & \\
	\;\vdots & & & \cdots & \\
	(S^-_j)^M & C_{M}^{1}(S^-_j)^{M-1} & C_{M}^{2}(S^-_j)^{M-2} & \cdots & \mathds{1}
\end{pmatrix*}\,.
\end{equation*}

\subsubsection{2. The ${\rm SU}(2s+1)$ ferromagnetic model}

For the ${\rm SU}(2s+1)$ ferromagnetic model,  a MPO  representation for $F_\alpha^{\,\,M_\alpha}$ ($\alpha=1,\;2,\;...,\;2s$) is  adapted from  that for a generic operator, developed in Ref.~~\cite{smyuping}, to the lowering operator $F_\alpha$ of ${\rm SU}(2s+1)$.
Hence, we have
\begin{equation}
	F_\alpha^{\,\,M_\alpha}= W_{[1]}^\alpha\cdots W^\alpha\cdots W_{[L]}^\alpha \,,
	\label{lo2}
\end{equation}
where the two matrices $W_1^\alpha$ and $W_L^\alpha$ at the two ends are
\begin{equation*}
	W_{[1]}^\alpha=
	\begin{pmatrix}
		F_{\alpha,j}^{\,\,M_\alpha}\; &
		C_{M_\alpha}^{1}F_{\alpha,j}^{\,\,M_\alpha-1}\; &
		C_{M_\alpha}^{2}F_{\alpha,j}^{\,\,M_\alpha-2}\; &
		\cdots &
		\mathds{1}
	\end{pmatrix}\,,
\end{equation*}
and
\begin{equation*}
	W_{[L]}^\alpha=\begin{pmatrix}
		\mathds{1} \\
		F_{\alpha,j} \\
		F_{\alpha,j}^{\,\,2} \\
		\;\vdots \\
		F_{\alpha,j}^{\,\,M_\alpha}
	\end{pmatrix}\,,
\end{equation*}
and  the bulk matrices
at the lattice sites $j=2,\;\ldots,\; L-1$ are identical, denoted as $W^\alpha$, which takes the form
\begin{equation*}
	W^\alpha=
	\begin{pmatrix*}[l]
		\,\mathds{1} & & & \vspace{5pt} \\
		F_{\alpha,j}& \;\mathds{1} & & & \vspace{5pt} \\
		F_{\alpha,j}^{\,\,2} & C_{2}^{1}F_{\alpha,j} & \;\mathds{1} & & \\
		\;\vdots & & & \cdots & \\
		F_{\alpha,j}^{\,\,M_\alpha} & C_{M_\alpha}^{1}F_{\alpha,j}^{\,\,M_\alpha-1} & C_{M_\alpha}^{2}F_{\alpha,j}^{\,\,M_\alpha-2} & \cdots & \mathds{1}
	\end{pmatrix*}\,.
\end{equation*}

\subsubsection{3. The staggered ${\rm SU}(3)$ spin-1 ferromagnetic biquadratic model}

For the staggered ${\rm SU}(3)$ spin-1 ferromagnetic biquadratic model,  $F_{\alpha,j}$ are different, depending on whether or not $j$ is even or odd. Hence, we introduce $ F_{\alpha,o}$ for odd sites and $F_{\alpha,e}$ for even sites.
A MPO representation for $F_\alpha^{\,\,M_\alpha}$  ($\alpha=1,\;2,\;3$) is  adapted from  that for a generic operator, developed in Ref.\\cite{smyuping}, to the lowering operator $F_\alpha$ of the staggered ${\rm SU}(3)$. Hence, we have
\begin{equation}
	F_\alpha^{\,\,M_\alpha}=W_{[1]}^\alpha\cdots W_e^\alpha W_o^\alpha\cdots W_{[L]}^\alpha \,,
	\label{lo3}
\end{equation}
where  the two matrices $W_1^\alpha$ and $W_L^\alpha$ at the two ends are
\begin{equation*}
	W_{[1]}^\alpha=
	\begin{pmatrix}
		F_{\alpha,o}^{\,\,M_\alpha}\; &
		C_{M_\alpha}^{1}F_{\alpha,o}^{\,\,M_\alpha-1}\; &
		C_{M_\alpha}^{2}F_{\alpha,o}^{\,\,M_\alpha-2}\; &
		\cdots &
		\mathds{1}
	\end{pmatrix}\,,
\end{equation*}
and
\begin{equation*}
	W_{[L]}^\alpha=\begin{pmatrix}
		\mathds{1} \\
		F_{\alpha,e} \\
		F_{\alpha,e}^{\,\,2} \\
		\;\vdots \\
		F_{\alpha,e}^{\,\,M_\alpha}
	\end{pmatrix}\,,
\end{equation*}
and the bulk matrices
at even sites are identical, denoted as $W_e^\alpha$, which takes the form
\begin{equation*}
	W_e^\alpha=
	\begin{pmatrix*}[l]
		\,\mathds{1} & & & \vspace{5pt} \\
		F_{\alpha,e}& \;\mathds{1} & & & \vspace{5pt} \\
		F_{\alpha,e}^{\,\,2} & C_{2}^{1}F_{\alpha,e} & \;\mathds{1} & & \\
		\;\vdots & & & \cdots & \\
		F_{\alpha,e}^{\,\,M_\alpha} & C_{M_\alpha}^{1}F_{\alpha,e}^{\,\,M_\alpha-1} & C_{M_\alpha}^{2}F_{\alpha,e}^{\,\,M_\alpha-2} & \cdots & \mathds{1}
	\end{pmatrix*}\,,
\end{equation*}
and the bulk matrices $W_o^\alpha$ at odd sites are identical, denoted as $W_o^\alpha$, which takes the form
\begin{equation*}
	W_o^\alpha=
	\begin{pmatrix*}[l]
		\,\mathds{1} & & & \vspace{5pt} \\
		F_{\alpha,o}& \;\mathds{1} & & & \vspace{5pt} \\
		F_{\alpha,o}^{\,\,2} & C_{2}^{1}F_{\alpha,o} & \;\mathds{1} & & \\
		\;\vdots & & & \cdots & \\
		F_{\alpha,o}^{\,\,M_\alpha} & C_{M_\alpha}^{1}F_{\alpha,o}^{\,\,M_\alpha-1} & C_{M_\alpha}^{2}F_{\alpha,o}^{\,\,M_\alpha-2} & \cdots & \mathds{1}
	\end{pmatrix*}\,.
\end{equation*}

This procedure offers a powerful means to construct an exact MPS representation for scale-invariant states in one-dimensional quantum many-body systems.
We remark that the bond dimension for such a MPS representation grows {\it linearly} with $L$.

\subsection{D: Exact matrix product state representations}

According to our prescription, a MPS representation for highly degenerate ground states results from contracting a MPO representation for the power of a lowering operator with a factorized state $|\psi_0\rangle$ or $|\psi_0\rangle_q$ representing the highest weight state or a generalized highest weight state with the period $q$, respectively.

\subsubsection{1. The ${\rm SU}(2)$ spin-$s$ ferromagnetic Heisenberg model}

An exact MPS representation for the highly degenerate ground states $|L,M\rangle=1/Z(L,M)(S_-)^{M}|\psi_0\rangle$, with the highest weight state $|\psi_0\rangle=\otimes_{l=1}^L\{s\}_l\rangle$,  follows from contracting the MPO representation for the power of the lowering operator $(S_-)^{M}$ in  Eq.(\ref{lo1}) with the MPS representation for the highest weight state $|\psi_0\rangle$, as specified in Eq.(\ref{hwsmps1}).
Hence, the highly degenerate ground states $|L,M\rangle=1/Z(L,M)(S_-)^{M}|\psi_0\rangle$ admit an exact MPS representation,
\begin{equation}
	|L,M\rangle = \sum_{s_1,\ldots,s_j,\ldots,s_L}  (B_{1}^{s_1}\cdots B^{s_j}\cdots B_{L}^{s_L})|s_1\cdots s_j\cdots s_L\rangle,
	\label{su2spinslmsm}
\end{equation}
where the two vectors $B_{1}^{s_1}$ and $B_{L}^{s_L}$ at the two ends take the form
\begin{equation*}
	B_{1}^{s_1}=\sum_{t}(W_{[1]})^{s_1t} A^{t},	
\end{equation*}
\begin{equation*}
	B_{L}^{s_L}=\sum_{t}(W_{[L]})^{s_Lt} A^{t},	
\end{equation*}
and the matrices $B^{s_j}$ at the lattice sites $j=2,\;\ldots, \;L-1$ are identical, which take the form
\begin{equation*}
	B^{s_j}=\sum_{t}W^{s_jt} A^{t}.
\end{equation*}
Thus, we are led to the explicit expressions for an exact MPS representation, presented in the main text (cf. Eq.(\ref{su2spinslm}) and below).

\subsubsection{2: The ${\rm SU}(2s+1)$ ferromagnetic model}

An exact MPS representation for the highly degenerate ground states $|L,M_1,\ldots,M_{2s}\rangle
={1}/{Z(L,M_1,\ldots,M_{2s})}\prod_{\alpha=1}^{2s}F_\alpha^{\,\,M_\alpha}|\psi_0\rangle$,, with the highest weight state $|\psi_0\rangle=\otimes_{l=1}^L\{s\}_l\rangle$,  follows from contracting the MPO representation for the power of the lowering operator $F_\alpha^{\,\,M_\alpha}$ ($\alpha=1$, $2$, \ldots, $2s$), in  Eq.(\ref{lo2}) with the MPS representation for the highest weight state $|\psi_0\rangle$, as specified in Eq.(\ref{hwsmps2}).
Hence, the highly degenerate ground states
$	|L,M_1,\ldots,M_{2s}\rangle
={1}/{Z(L,M_1,\ldots,M_{2s})}\prod_{\alpha=1}^{2s}F_\alpha^{\,\,M_\alpha}|\psi_0\rangle$ admit an exact MPS representation,
\begin{equation}
	|L,M_1,\ldots,M_{2s}\rangle = \sum_{s_1,\ldots,s_j,\ldots,s_L}  (B_1^{s_1}\cdots B^{s_j}\cdots B_L^{s_L})|s_1\cdots s_j\cdots s_L\rangle,
	\label{su2sp1lmsm}
\end{equation}
where the two vectors $B_1^{s_1}$ and $B_L^{s_L}$ at the two ends take the form
\begin{equation*}
	B_1^{s_1}=\sum_{t_{1},\ldots,t_{\alpha},\ldots,t_{2s}}
	(W_{[1]}^{1})^{s_1t_1}\cdots(W_{[1]}^{\alpha})^{t_{\alpha-1}t_{\alpha}}\cdots (W_{[1]}^{2s})^{t_{2s-1}t_{2s}}A^{t_{2s}},
\end{equation*}

\begin{equation*}
	B_L^{s_L}=\sum_{t_{1},\ldots,t_{\alpha},\ldots,t_{2s}}
	(W_{[L]}^{1})^{s_1t_1}\cdots(W_{[L]}^{\alpha})^{t_{\alpha-1}t_{\alpha}}\cdots	(W_{[L]}^{2s})^{t_{2s-1}t_{2s}}A^{t_{2s}},
\end{equation*}
and the matrices $B^{s_j}$ at the lattice sites $j=2,\;\ldots, \;L-1$ are identical, which take the form
\begin{equation*}
	B^{s_j}=\sum_{t_{1},\ldots,t_{\alpha},\ldots,t_{2s}}
	(W^{1})^{s_jt_1}\cdots(W^{\alpha})^{t_{\alpha-1}t_{\alpha}}\cdots	(W^{2s})^{t_{2s-1}t_{2s}}A^{t_{2s}}.
\end{equation*}
Thus, we are led to the explicit expressions for an exact MPS representation, presented in the main text (cf. Eq.(\ref{su2sp1lm}) and below).

\subsubsection{3: The staggered ${\rm SU}(3)$ spin-1 ferromagnetic biquadratic model}

An exact MPS representation for the highly degenerate ground states $|L,M_1,M_2\rangle_2=1/Z_2(L,M_1,M_2)F_1^{M_1}F_2^{M_2}|\psi_0\rangle$, with the highest weight state $|\psi_0\rangle=\otimes_{l=1}^L\{s\}_l\rangle$, follows from contracting the MPO representation for the power of the lowering operator $F_\alpha^{\,\,M_\alpha}$ ($\alpha=1$, $2$), in  Eq.(\ref{lo3}) with the MPS representation for the highest weight state $|\psi_0\rangle$, as specified in Eq.(\ref{hwsmps3}).
Hence,  the highly degenerate ground states $|L,M_1,M_2\rangle_2=1/Z_2(L,M_1,M_2)F_1^{M_1}F_2^{M_2}|\psi_0\rangle$ admit an exact MPS representation,
\begin{align}
	|L,M_1,M_2\rangle_2 = &\sum_{s_1,\ldots,s_L}  (B_1^{s_1}\cdots B_{[1]}^{s_{2j-1}}B_{[2]}^{s_{2j}}\cdots B_L^{s_L})\nonumber \\
	&|s_1\cdots s_{2j-1}s_{2j}\cdots s_L\rangle,
	\label{lm1m2}
\end{align}
where the two vectors $B_{1}^{s_1}$ and $B_{L}^{s_L}$ at the two ends take the form
\begin{equation*}
	B_{1}^{s_1}=\sum_{t_{1},t_{2}}
	(W_{[1]}^{1})^{s_1t_1}	(W_{[1]}^{2})^{t_{1}t_{2}}A^{t_{2}},
\end{equation*}
\begin{equation*}
	B_{L}^{s_L}=\sum_{t_{1},t_{2}}
	(W_{[L]}^{1})^{s_Lt_1}	(W_{[L]}^{2})^{t_{1}t_{2}}A^{t_{2}},
\end{equation*}
and the matrices $B_{[1]}^{s_{2j-1}}$ at the $(2j-1)$-th lattice sites, with $j=2,\;\ldots, \;L/2$, take the form
\begin{equation*}
	B_{[1]}^{s_{2j-1}}=\sum_{t_{1}t_{2}}
	(W_{o}^{1})^{s_{2j-1}t_1}(W_{o}^{2})^{t_{1}t_{2}}A^{t_{2}},
\end{equation*}
and the matrices $B_{[2]}^{s_{2j}}$ at the $2j$-th lattice sites, with $j=1,\;\ldots, \;L/2-1$, take the form
\begin{equation*}
	B_{[2]}^{s_{2j}}=\sum_{t_{1}t_{2}}
	(W_{e}^{1})^{s_{2j}t_1}(W_{e}^{2})^{t_{1}t_{2}}A^{t_{2}}.
\end{equation*}
Thus, we are led to the explicit expressions for an exact MPS representation, presented in the main text (cf. Eq.(\ref{lm1m2}) and below).

An exact MPS representation for the highly degenerate ground states $|L,M_2,M_3\rangle_4=1/Z_4(L,M_2,M_3)F_2^{M_2}F_3^{M_3}|\psi_0\rangle_4$, with the generalized highest weight state $|\psi_0\rangle_4=|\otimes_{l=1}^{L/4}\{1110\}_l\rangle$, follows from contracting the MPO representation for the power of the lowering operator $F_\alpha^{\,\,M_\alpha}$ ($\alpha=2$, $3$), in  Eq.(\ref{lo3}) with the MPS representation for the generalized highest weight state $|\psi_0\rangle_4$, as specified in Eq.(\ref{hwsmps4}).
Hence,  the highly degenerate ground states $|L,M_2,M_3\rangle_4=1/Z_4(L,M_2,M_3)F_2^{M_2}F_3^{M_3}|\psi_0\rangle_4$ admit an exact MPS representation,
\begin{align}
	&|L,M_2,M_3\rangle_4 = \sum_{s_1,\ldots,s_L} (B_{1}^{s_1}B_{[2]}^{s_2}B_{[3]}^{s_3}B_{[4]}^{s_4}\cdots \nonumber \\
	& B_{[1]}^{s_{4m-3}}B_{[2]}^{s_{4m-2}}B_{[3]}^{s_{4m-1}}B_{[4]}^{s_{4m}} \cdots B_{[1]}^{s_{L-3}}B_{[2]}^{s_{L-2}}B_{[3]}^{s_{L-1}}B_L^{s_L}) \nonumber \\
	&|s_1s_2s_3s_4\cdots s_{4m-3}s_{4m-2}s_{4m-1}s_{4j}\cdots s_{L-3}s_{L-2}s_{L-1}s_L\rangle,
	\label{Lm2m3g}
\end{align}
where the two vectors $B_1^{s_1}$ and $B_L^{s_L}$ at the two ends take the form
\begin{equation*}
	B_1^{s_1}=\sum_{t_{1},t_{2}}
	(W_{[1]}^{2})^{s_1t_1}	(W_{[1]}^{3})^{t_{1}t_{2}}A^{t_{2}},
\end{equation*}
\begin{equation*}
	B_L^{s_L}=\sum_{t_{1},t_{2}}
	(W_{[L]}^{2})^{s_Lt_1}	(W_{[L]}^{3})^{t_{1}t_{2}}A^{t_{2}},
\end{equation*}
and the matrices $B_{[1]}^{s_{4m-3}}$ at the $(4m-3)$-th lattice sites, with $m=2,3,\;\ldots, \;L/4$, take the form
\begin{equation*}
	B_{[1]}^{s_{4m-3}}=\sum_{t_{1}t_{2}}
	(W_{o}^{1})^{s_{4m-3}t_1}(W_{o}^{2})^{t_{1}t_{2}}A^{t_{2}},
\end{equation*}
the matrices $B_{[2]}^{s_{4m-2}}$ at the $(4m-2)-th$ lattice sites, with $m=1,2,\;\ldots, \;L/4$, take the form
\begin{equation*}
	B_{[2]}^{s_{4m-2}}=\sum_{t_{1}t_{2}}
	(W_{e}^{1})^{s_{4m-2}t_1}(W_{e}^{2})^{t_{1}t_{2}}A^{t_{2}},
\end{equation*}
the matrices $B_{[3]}^{s_{4m-1}}$ at the $(4m-1$)-th lattice sites, with  $m=1,2,\;\ldots, \;L/4$, take the form
\begin{equation*}
	B_{[3]}^{s_{4m-1}}=\sum_{t_{1}t_{2}}
	(W_{o}^{1})^{s_{4m-1}t_1}(W_{o}^{2})^{t_{1}t_{2}}A^{t_{2}},
\end{equation*}
and the matrices $B_0^{s_{4m}}$ at the $4m$-th lattice sites, with $m=1,2,\;\ldots, \;L/4-1$, take the form
\begin{equation*}
	B_{[4]}^{s_{4m}}=\sum_{t_{1}t_{2}}
	(W_{e}^{1})^{s_{4m}t_1}(W_{e}^{2})^{t_{1}t_{2}}(A_0)^{t_{2}}.
\end{equation*}
Thus, we are led to the explicit expressions for an exact MPS representation, presented in the main text (cf. Eq.(\ref{Lm2m3g}) and below).

\subsection{E. Finite system-size scaling of the entanglement entropy and the number of type-B Goldstone modes $N_B$}

For the three illustrative models, $N_B$ is extracted from performing a  universal finite system-size scaling analysis of the entanglement entropy $S_f(L,n)$~\cite{smfinitesize} (cf. Eq.(\ref{slnf}) in the main text). Here, we have treated both $N_B$ and $S_{\!\!f0}$ as the fitting parameters, and exploited a deviation of  $N_B$ from its exact values as a measure of the accuracy for our numerical data, evaluated from an exact MPS representation for highly degenerate ground states.

\begin{table}
	\centering
	\caption{  $N_B$ and $S_{\!\!f0}$ extracted from  performing a universal finite system-size scaling analysis of the entanglement entropy $S_f(L,n)$ for the ${\rm SU}(2)$ spin-$s$ ferromagnetic Heisenberg model: (a) $s=1/2$, with $f$ chosen to be $f=3/10$, $1/2$ and $3/4$. (b) $s=1$, with $f$ chosen to be $f=2/5$, $3/5$ and $1$, respectively. Here, $L=100$ and $n$ ranges from 10 to 50.
	}
	\vspace{2mm}
	\label{tab1}
	\addtolength{\tabcolsep}{5pt}
	\begin{tabular}{|c|c|c|c|c|}
		\hline
		$s$ &   &$f$ & $N_B$ & $S_{\!\!f0}$\\
		\hline
		& & $3/10$ &1.012&0.899\\
		$\frac{1}{2}$ & $|L,M\rangle$ & $1/2$ &1.000 & 1.052\\
		& & $3/4$ &1.023&    0.795\\
		\hline
		& &2/5& 1.016&1.194\\
		1 &  $|L,M\rangle$& $3/5$ &1.004 &   1.414\\
		& &1 &1.000 & 1.550\\
		\hline
	\end{tabular}
\end{table}

\begin{table}[htbp]
	\centering
	\caption{  $N_B$ and $S_{\!\!f0}$ extracted from  performing a universal finite system-size scaling analysis of the entanglement entropy $S_f(L,n)$ for the  ${\rm SU}(2s+1)$ ferromagnetic model: (a) For $s=1$, with $f_1=1/5$ and $f_2=1/5$, $f_1=1/5$ and $f_2=1/4$, and $f_1=3/10$ and $f_2=1/4$, respectively. (b) For  $s=3/2$, with  $f_1=1/6$, $f_2=1/6$ and $f_3=1/6$, $f_1=1/6$, $f_2=1/5$ and $f_3=1/5$, and $f_1=1/5$, $f_2=1/5$ and $f_3=1/5$, respectively. Here,  $L=100$ and $n$ ranges from 10 to 50 for the ${\rm SU(3)}$ ferromagnetic model and $L=80$ and $n$ ranges from 20 to 40 for the ${\rm SU(4)}$ ferromagnetic model, respectively.
	}
	\vspace{2mm}
	\label{tab2}
	\begin{tabular}{|c|c|c|c|}
		\hline
		& $f$ & $N_B$ & $S_{\!\!f0}$\\
		\hline
		&$(1/5,\;1/5)$ &2.050&1.303\\
		$|L,M_1, M_2\rangle$  &$(1/5,\;1/4)$  &2.040& 1.424\\
		& $(3/10,\;1/4)$  &2.025&  1.605\\
		\hline
		& $(1/6,1/6,1/6)$ &3.107&1.718\\
		$|L, M_1, M_2,M_3\rangle$  &$(1/6,1/5,1/5)$ &3.099 &   1.787\\
		& $(1/5,1/5,1/5)$ &3.094 & 1.821\\
		\hline
	\end{tabular}
\end{table}

\begin{table}
	\centering
	\caption{  $N_B$ and $S_{\!\!f0}$  extracted from  performing a  universal finite system-size scaling analysis of the entanglement entropy $S_f(L,n)$  for the staggered ${\rm SU}(3)$ spin-1 ferromagnetic biquadratic model: (a) For $|L,M_1,M_2\rangle_2$, with the fillings $f_1=3/20$ and $f_2=1/5$, $f_1=1/5$ and $f_2=1/5$, and $f_1=1/4$ and $f_2=3/10$, respectively. (b)  For $|L,M_1^*,M_2^*\rangle_4$, with the fillings $f_1^*=3/20$ and $f_2^*=3/20$, $f_1^*=3/20$ and $f_2^*=1/5$ and $f_1^*=1/10$ and $f_2^*=1/4$, respectively. In both cases, $L=100$, and $n$ ranges from 20 to 50.
	}
	\vspace{3mm}
	\label{tab3}
	\begin{tabular}{|c|c|c|c|}
		\hline
		& $f$ & $N_B$ & $S_{\!\!f0}$\\
		\hline
		$|L,M_1,M_2\rangle_2$  & $(3/20,\;1/5)$ &2.045&1.031\\
		& $(1/5,\;1/5)$  &2.029& 1.151\\
		& $(1/4,\;3/10)$  &2.016& 1.342\\
		\hline
		$|L,M_1^*,M_2^*\rangle_4$ & $(3/20,3/20)$ &2.058&0.395\\
		&$(3/20,1/5)$ &2.038 & 0.562\\
		& $(1/10,1/4)$ &2.032 & 0.679\\
		\hline
	\end{tabular}
\end{table}

As shown in Table~\ref{tab1}, $N_B$ and $S_{\!\!f0}$ are extracted from  performing a finite system-size universal scaling analysis of the entanglement entropy $S_f(L,n)$ for the ${\rm SU}(2)$ spin-$s$ ferromagnetic Heisenberg model: (a) $s=1/2$, with $f$ chosen to be $f=3/10$, $1/2$ and $3/4$. (b) $s=1$, with $f$ chosen to be $f=2/5$, $3/5$ and $1$, respectively. Here, $L=100$ and $n$ ranges from 10 to 50.

As shown in Table~\ref{tab2}, $N_B$ and $S_{\!\!f0}$ are extracted from  performing a  universal finite system-size scaling analysis of the entanglement entropy $S_f(L,n)$ for the  ${\rm SU}(2s+1)$ ferromagnetic model: (a) For $s=1$, with $f_1=1/5$ and $f_2=1/5$, $f_1=1/5$ and $f_2=1/4$, and $f_1=3/10$ and $f_2=1/4$, respectively. (b) For  $s=3/2$, with  $f_1=1/6$, $f_2=1/6$ and $f_3=1/6$, $f_1=1/6$, $f_2=1/5$ and $f_3=1/5$, and $f_1=1/5$, $f_2=1/5$ and $f_3=1/5$, respectively. Here,  $L=100$ and $n$ ranges from 10 to 50 for the ${\rm SU(3)}$ ferromagnetic model and $L=80$ and $n$ ranges from 20 to 40 for the ${\rm SU(4)}$ ferromagnetic model, respectively.

As shown in Table~\ref{tab3},  $N_B$ and $S_{\!\!f0}$ are extracted from  performing a universal finite system-size scaling analysis of the entanglement entropy $S_f(L,n)$  for the staggered ${\rm SU}(3)$ spin-1 ferromagnetic biquadratic model: (a) For $|L,M_1,M_2\rangle_2$, with $f_1=3/20$ and $f_2=1/5$, $f_1=1/5$ and $f_2=1/5$, and $f_1=1/4$ and $f_2=3/10$, respectively. (b)  For $|L,M_1^*,M_2^*\rangle_4$, with $f_1^*=3/20$ and $f_2^*=3/20$, $f_1^*=3/20$ and $f_2^*=1/5$ and $f_1^*=1/10$ and $f_2^*=1/4$, respectively. In both cases, $L=100$, and $n$ ranges from 20 to 50.

A conclusion to be drawn from our discussions is that the number of type-B GMs extracted from performing a finite system-size universal scaling analysis of the entanglement entropy $S_f(L,n)$ is consistent with that from the counting rule of GMs~\cite{smwatanabe}.

\section{Acknowledgements}
We are grateful to Murray Batchelor and John Fjaerestad  for helpful discussions.
I.P.M. acknowledges funding from the National Science and Technology Council (NSTC) Grant No. 122-2811-M-007-044.

\end{document}